\documentclass[a4paper, oneside, twocolumn, notitlepage, 10pt]{extarticle_ecoc}
\usepackage{ecoc}

\usepackage{filecontents}

\begin{document}
\selectlanguage{english}    


\title{Experimental Comparison of Average-Power Constrained and Peak-Power Constrained 64QAM under Optimal Clipping in 400Gbps Unamplified Coherent Links}%


\author{
    Wing-Chau Ng\textsuperscript{(1)} and Chuandong Li\textsuperscript{(1)} 
}

\maketitle                  


\begin{strip}
    \begin{author_descr}

        \textsuperscript{(1)}Huawei Technologies Co., Ltd., Ottawa, ON K2K 3J1, Canada,
        \textcolor{blue}{\uline{wing.chau.ng@huawei.com}}

    \end{author_descr}
\end{strip}

\renewcommand\footnotemark{}
\renewcommand\footnoterule{}

\begin{strip}
    \begin{ecoc_abstract}
        We experimentally demonstrated an end-to-end link budget optimization over clipping in 400Gbps unamplified links, showing that the clipped MB distribution outperforms the peak-power constrained 64QAM by 1dB link budget. \textcopyright2024 The Author(s)
    \end{ecoc_abstract}
\end{strip}

\section{Introduction}
While chromatic dispersion and  four-wave mixing limit the reach of the intensity-modulation direct detection (IM-DD) systems for the future 3.2T data center interconnects\cite{example:conference0}, optical circuit switching further incurs higher link loss\cite{example:Apollo}. Coherent solutions would become necessary for both unamplified links and short reach, i.e., bundling LR (10km) with ZR (80-120km)\cite{example:OFC2024Ciena, example:OFC2024CienaE}. Higher spectral efficiency (SE) coherent modulation formats suffer more from modulator loss and driver nonlinearity, which limits transmitter (Tx) power and thus reduces the link loss budget\cite{example:ecoc2024imec}. While using optical booster or higher laser power is required for higher Baud rates, signaling or shaping design is also indispensable to further maximize the Tx output power to compensate the link loss \cite{example:PPCimdd2021, example:NokiaIMDD}.  
 
The additive white Gaussian noise (AWGN) “capacity” is defined given an average power constraint (APC). Maxwell-Boltzmann (MB) distribution is a well-known capacity-achieving, rate-adaptive option \cite{example:Buchali2016}. Recently, a peak-power constraint (PPC) has become a popular concept in amplifier-less scenarios such as in 400ZR scenarios \cite{example:ECOC2023PPC}.  However, once the APC is “inactive” \cite{example:Bocherer2011}, the concept of “capacity-achieving” should not be applied, since increasing power arbitrarily enhances mutual information (MI) in an AWGN channel. Therefore, PPC-64QAM\cite{example:ECOC2023PPC, example:ofc2024nokiaAI} is not capacity-achieving, but only “MI-maximizing” under PPC. Otherwise, a new “capacity under PPC” should be defined and derived first.
\begin{figure}[t!]
	\centering
	\includegraphics[height=3.7cm]{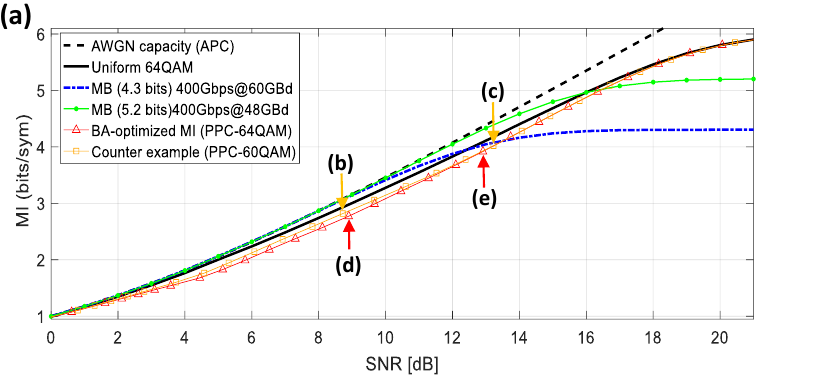}
	\includegraphics[height=5cm]{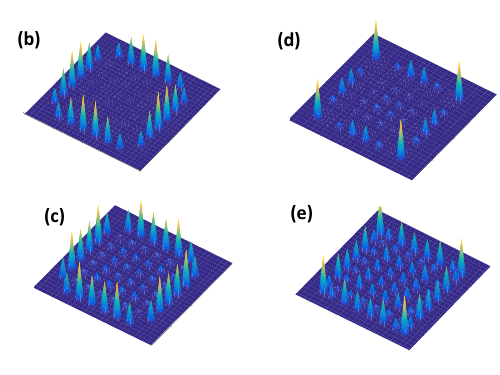}
	\caption{Achievable information rates versus SNR for APC (MB) and BA-optimized PPC shaping. (b,c) A counterexample with higher MI under PPC. (d,e) PPC-64QAM previously proposed\cite{example:ECOC2023PPC}.}
	\label{fig:chanCap}
\end{figure}
 
Numerous PPC-related works   \cite{example:ECOC2023PPC, example:PPCimdd2021, example:NokiaIMDD, example:clipping2021, example:ofc2024nokiaAI, example:ofc2023rmb} share a common goal: Given a finite modulator depth (or waveform’s peak power), how can the Tx power be made sufficiently high to survive below the forward error correction (FEC) threshold at receiver (Rx)? Is the already-existing MB distribution still useful in the future 1600ZR\cite{example:OFC2024Ciena, example:OFC2024CienaE}?

The desirable waveforms should have low peak-to-average power ratio (PAPR). To control the PAPR and the waveform peak magnitude, aggressively clipping MB distribution had been demonstrated to enhance power gain \cite{example:clipping2021}. For bandwidth-limited channels, clipping optimization becomes critical as S21 pre-compensation enhances PAPR substantially \cite{example:NokiaIMDD, example:ofc2024nokiaAI, example:ecoc2023DPE}. Thus, clipping must be another key parameter in ZR scenarios. Currently, factory calibration optimizes clipping based on the back-to-back (B2B) required signal-to-noise ratio (RSNR), without considering the end-to-end (E2E) link loss. For instance, a recent work \cite{example:ECOC2023PPC} does not provide a fair comparison between APC- and PPC-64QAM since MB is highly sensitive to clipping. 

In this regard, this work provides a fair com-parison of APC- and PPC-64QAM under optimal clipping for an E2E performance. First, we show that PPC-64QAM is not capacity-achieving, with a counterexample. Next, we explain why the conventional B2B clipping optimization works for PPC-64QAM but not for MB. Then, we analyse and discuss our experimental results on optimal clipping for the E2E link budget. Finally we conclude this work.
\section{PPC is not capacity achieving}
Since the conventional channel coding and the concept of capacity-achieving relies on SNR rather than peak SNR \cite{example:PPCimdd2021, example:NokiaIMDD, example:ofc2024nokiaAI}, it is important to investigate if PPC-optimized input probability mass function (PMF) achieves the AWGN capacity (as claimed previously \cite{example:ECOC2023PPC}) by plotting the MI over SNR. 
\begin{figure}[t!]
	\centering
	\includegraphics[height=4.2cm]{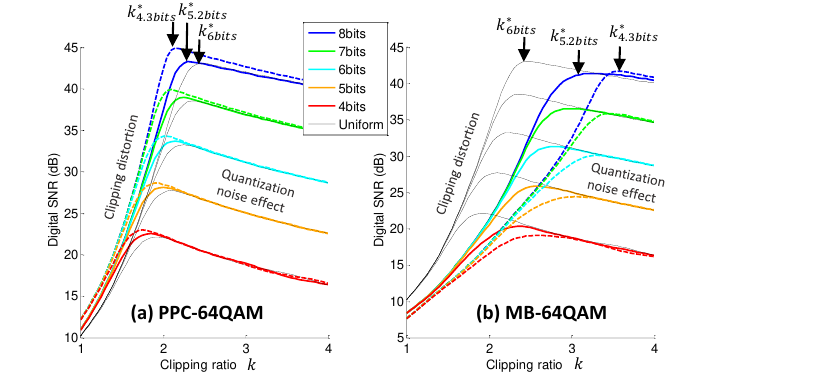}
	\caption{ Conventional digital SNR-based clipping optimization for (a) PPC-64QAM (b) MB-64QAM implemented for 400 Gbps system. Solid: 60 GBaud (4.3 bits/sym), dashed: 48 GBaud (5.2 bits/sym).}
	\label{fig:SNRopt}	
\end{figure}
Fig.~\ref{fig:chanCap}a shows the comparison of achievable information rates (AIR) of two MB-64QAMs for 400Gbps at 48 Gbaud (green dot) and 60 GBaud (blue dash-dotted), uniformly distributed (UD) 64QAM (solid black) and Blahut-Arimoto (BA)-optimized PPC-64QAM (red triangle). Caution must be taken on the true SNR since the power of optimized PPC-64QAM’s constellation keeps changing at different noise variances with respective to the original constellation power, as explained in \cite{example:UofTawgn}. The so-called “PPC” is merely an implicit constraint (i.e., given a constellation, peak symbols are fixed) of an unconstrained MI maximization problem solved by the unconstrained BA algorithm. The AIR of PPC-64QAM is lower than the AWGN capacity and even that of the UD-64QAM. Morever, there exists, as a counterexample, another PMF having a higher AIR under PPC, i.e., the BA-optimized “60QAM”, shown in orange in Figs.~\ref{fig:chanCap}a-c.  Thus, “capacity-achieving” is not appropriately used for PPC-64QAM\cite{example:ECOC2023PPC}.

\section{Drawbacks of B2B Clipping Optimization}
Conventionally, clipping is optimized based on the B2B digital SNR of the waveform loaded into the digital-to-analog converter (DAC) with its nominal number of converter bits, $n$ \cite{example:ecoc2018NEC, example:zChenTxOpt}. An ideal waveform, $x(t)$, with a symmetric distribution $f_x(x)$ with its original variance $\sigma_x^2$, is clipped at $ c= k \sigma_x$, where $k$  is defined as the clipping ratio in this work. The clipped waveform variance becomes $2\int_{c}^{\infty} c^2 f_x(x) dx  + 2 \int_{0}^{c} x^2 f_x(x) df$, while the clipping noise variance is $ 2 \int_{c}^{\infty} (x-c)^2 f_x(x) dx $. As $c$ is smaller than the half-scale swing, $A_{max} = 2^{(n-1)}$, the clipped signal (with noise together) is rescaled by $\frac{A_{max}}{c}$ to cover the full-scale swing to avoid quantization noise. Therefore, the SNR is
\linespread{1}
\begin{equation}\label{eq:eq1}
   SNR(k) \approx \frac{k^2 \sigma^2_x \int_{k\sigma_x}^{\infty} f_x(x)dx + \int_{0}^{k\sigma_x}  x^2  f_x(x)dx }{\frac{k^2 \sigma^2_x}{6 ( 2^{2n}) }  + \int_{k\sigma_x}^{\infty} (x-k\sigma_x)^2 f_x(x) dx  } 
\end{equation}
for any $ f_x(x)$. 
\begin{figure}[t!]
	\centering
	\includegraphics[height=4.1cm]{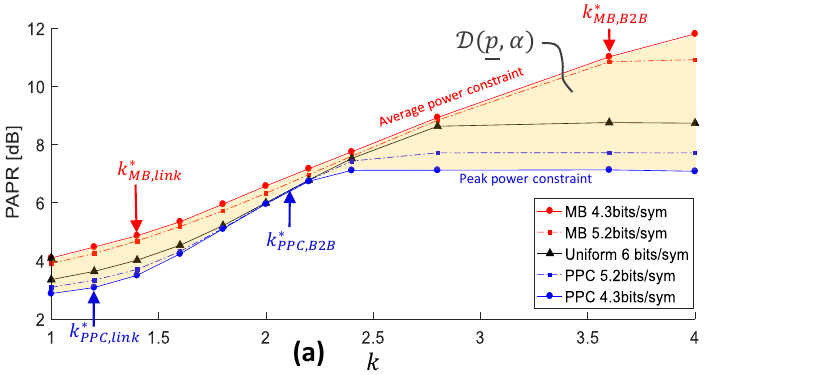}
	\includegraphics[height=4.1cm]{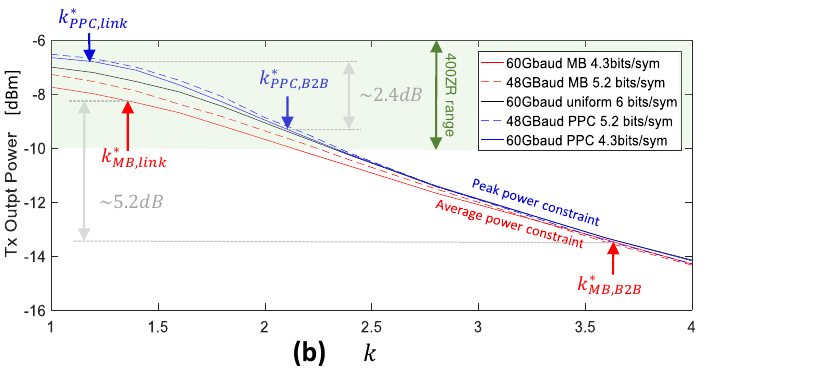}
	\caption{ (a) PAPR versus clipping ratio for MB- and APC-64QAM. (b) The measured Tx output power versus $k$ for 400Gbps at 60 GBaud and 48 Gbaud. $\alpha=0.2$.}
	\label{fig:papr}	
\end{figure}
Fig.~\ref{fig:SNRopt} shows the conventional clipping optimization for (a) PPC and (b) MB distribution, with solid lines for 5.2 bits/sym (60 Gbaud) and dashed lines for 4.3 bits/sym (48 Gbaud) to achieve 400Gbps, with respect to 6 bits/sym as a benchmark (black, dashed). Throughout this report, a root-raised cosine (RRC) pulse shape with a roll-off factor $\alpha=0.2$ is used as in \cite{example:ECOC2023PPC}. The concavity indicates a balance between clipping distortion and quantization noise. For PPC-64QAM, optimal $k^*$ is insensitive to the entropy range (4.3-6 bits/sym), but $k^*$ changes from 2.2 to 3.6 for MB-64QAM, reducing Tx output power from -10 dBm to -14 dBm (below 400ZR minimum power\cite{example:oif400G}), as shown in Fig.~\ref{fig:papr}b. Fig.~\ref{fig:papr}a compares the PAPR as a function of $k$ among various PMFs, with their B2B optimal $k_{MB,B2B}^*=3.6$ and $k_{PPC,B2B}^*=2.1$.  Fig.~\ref{fig:papr}b recorded the Tx output power as a function of $k$. The output powers at $k_{MB,B2B}^*$ and $k_{PPC,B2B}^*$ are -13.5 dBm and -9.2 dBm, i.e. penalize MB to avoid clipping distortion because of high PAPR. This explains why using the conventional B2B clipping optimization does not allow a fair link-loss comparison between APC- and MB-64QAM in ZR scenarios. However, as shown in Fig.~\ref{fig:papr}b, heavy clipping gives 5.2 dB and 2.4 dB extra power gain to MB and PPC, respectively. A fair comparison should include the balance between clipping distortion and power gain, to be discussed next.

\section{Link Budget Experiment}
\begin{figure*}[t]
	\centering
	\includegraphics[height=4.4cm]{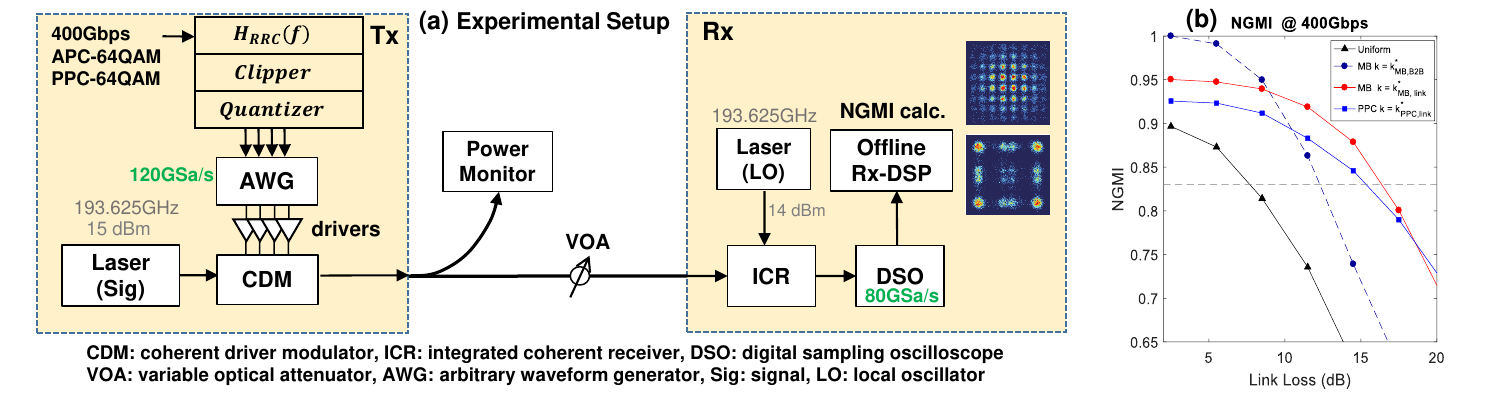}
	\caption{ Experimental setup. (b) Experimental performance in terms of NGMI for each input distribution at 400Gbps.}
	\label{fig:setup}
\end{figure*}

\begin{figure}[t!]
	\centering
	\includegraphics[height=4cm]{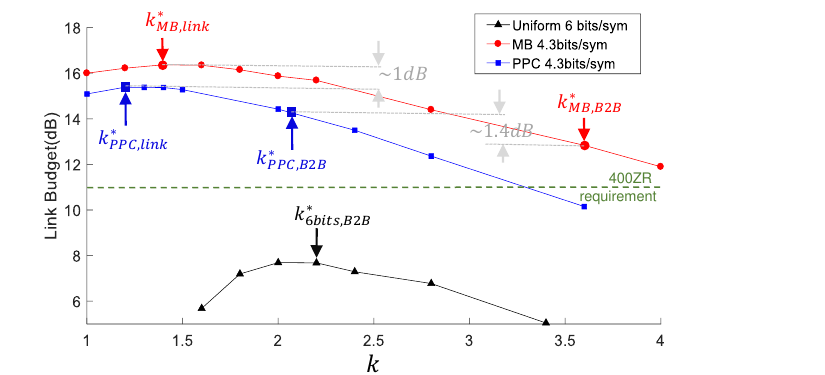}
	\caption{ Experimentally achieved link budget versus clipping ratio at 400Gbps (red and blue). $k^*$: optimal clipping ratios.}
	\label{fig:budget}	
\end{figure}

Fig.~\ref{fig:setup}a shows the experimental setup of a 400Gbps unamplified link. A 60-GBaud RRC pulse shaped ($\alpha=0.2$) 64QAM with various PMFs was generated by a 120-Gsa/s arbitary waveform generator. The driver peak voltage was set to 350 mV. The Tx and Rx lasers were operating at 193.625 GHz with power levels of 15 dBm and 14 dBm, respectively. The measured launch power was between -15 dBm and -7 dBm, shown in Fig.~\ref{fig:papr}b. A variable optical attenuator (VOA) emulated the link loss. At Rx, a coherent receiver down-converted the optical signal to electrical baseband, which was sampled by a digital sampling oscilloscope (DSO) with four 80-GSa/s lanes each having analog bandwidths of 33 GHz. Tx and Rx skews were first jointly calibrated using \cite{example:mimo8x4} and compensated. A single-stage 275-tap 4×8 MIMO equalizer \cite{example:mimo8x4} was used to compensate the Tx and Rx IQ crosstalk, intersymbol interference and polarization rotation. The link budget is defined as the maximum link loss under which the measured normalized generalized (NG) MI is above the FEC rate of 5/6, shown as the dashed level in Fig.~\ref{fig:setup}b, and reported in Fig.~\ref{fig:budget} for comparison. 

\begin{figure}[t!]
	\centering
	\includegraphics[height=5.2cm]{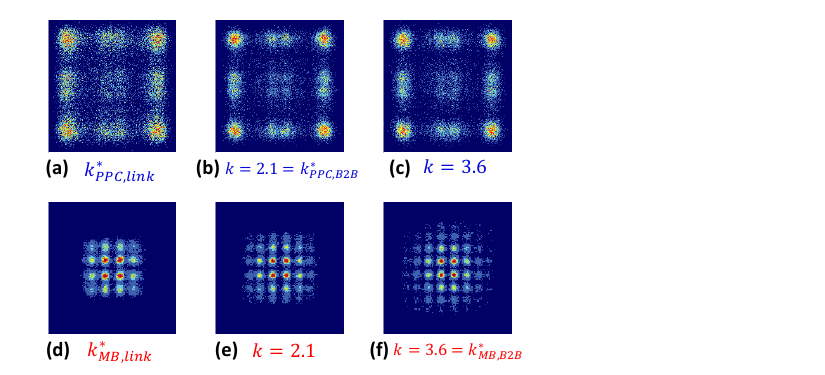}
	\caption{ B2B constellations of (a-c) clipped PPC-64QAM and (d-f) clipped MB-64QAM at various $k$ shown in Fig.~\ref{fig:budget}.}
	\label{fig:constel}	
\end{figure}
From Fig.~\ref{fig:budget}, MB-64QAM suffers from 1.4 dB budget loss at $k_{MB,B2B}^*$ compared to PPC-64QAM. With E2E optimal clipping, MB-64QAM outperforms PPC-64QAM by 1 dB at $k_{MB,link}^*$.  Uniform PMF (500Gbps) has the worst budget, but it serves as a benchmark only. Compared to \cite{example:ECOC2023PPC}, all curves have ~3 dB budget again, because our Tx and Rx laser power were 2 dB and 1 dB higher than the values shown in the previous work \cite{example:ECOC2023PPC}. At $k_{PPC,link}^*$  and $k_{MB,link}^*$ , shaped 9-QAM-like and 16-QAM-like constellation are found in Figs.~\ref{fig:constel}a and d, respectively. Fig.~\ref{fig:budget} shows that MB-64QAM clipped at $k = 2.1$ has a similar budget as PPC-64QAM at $k_{PPC,link}^*$, showing a shaped 36-QAM-like constellation, shown in Fig.~\ref{fig:constel}e.

Interestingly, in Fig.~\ref{fig:papr}a, the orange region, $ \mathcal{D}(\underline{p}, \alpha)$, the set of possible PAPR and $k$,  bounded between the APC (red) and the PPC (blue), has higher entropy but lower link budget. UD-64QAM always lies at the middle of $\mathcal{D}(\underline{p}, \alpha)$ and has the worst link budget. Closer to the lower bound (blue), clipping enhances link budget with small distortion on constellation, shown in Figs. \ref{fig:constel}a-c.  Closer to the upper bound (red), the mature MB shaping outperforms PPC-64QAM by 1dB, suggesting that the previous work\cite{example:ECOC2023PPC} does not optimize MB-64QAM well for a fair comparison.  Any $\underline{p}$ leading to $\mathcal{D}^c$ is not economical because higher device bandwidth is required to achieve the same rate.

\section{Conclusions}
In this work, we show that PPC-64QAM is not capacity-achieving. The conventional SNR-based clipping optimization works well for PPC-64QAM, but not for MB-64QAM. However, by considering clipping as a key variable for E2E optimization, MB-64QAM outperforms PPC-64QAM by 1 dB budget. Nevertheless, whether MB is the best option under clipping requires further research, e.g., using APC-BA algorithm in an AWGN channel with clipping for a specific pulse shape or waveform distribution. 

\clearpage

\defbibnote{myprenote}{%

}
\printbibliography[prenote=myprenote]
\end{document}